\documentclass[notitlepage]{revtex4-1}
\usepackage[margin=1in]{geometry}
\usepackage{amsmath}
\usepackage{amsthm}
\usepackage{amssymb}
\usepackage{graphicx}
\usepackage{subfig}
\usepackage{caption}
\usepackage{float}
\usepackage{mathrsfs}
\usepackage{hyperref}
\usepackage[usenames,dvipsnames]{xcolor}
\hypersetup{colorlinks=true,linkcolor=.,citecolor=blue,urlcolor=blue}

\theoremstyle{plain}

\theoremstyle{definition}

\DeclareMathOperator{\sgn}{sgn}

\renewcommand{\a}{\alpha}
\renewcommand{\b}{\beta}

\renewcommand{\d}{\delta}

\renewcommand{\th}{\theta}

\renewcommand{\l}{\lambda}
\renewcommand{\r}{\rho}
\newcommand{\s}{\sigma}

\newcommand\inv{^{-1}}
\newcommand{\la}{\langle}
\newcommand{\ra}{\rangle}
\renewcommand{\=}{\ =\ }

\newcommand{\pd}[2]{\frac{\partial{#1}}{\partial{#2}}}

\newcommand{\pds}[2]{\frac{\partial^2{#1}}{\partial{#2}^2}}

\newcommand{\be}{\begin{equation}}
\newcommand{\ee}{\end{equation}}
\newcommand{\bea}{\begin{eqnarray}}
\newcommand{\eea}{\end{eqnarray}}
\newcommand{\beas}{\begin{eqnarray*}}
\newcommand{\eeas}{\end{eqnarray*}}
\newcommand{\lint}{\int^b_a}
\newcommand{\bl}{\Lambda}
\newcommand{\wt}{\widetilde}

\makeatletter
\def\blfootnote{\gdef\@thefnmark{}\@footnotetext}
\makeatother

\usepackage[dvipsnames]{xcolor}

\usepackage{fancyhdr}
\begin{document}

\title{Constructing Solutions to Two-Way Diffusion Problems}



\author{Caleb G. Wagner}
\email{c.g.wagner23@gmail.com}
\affiliation{Martin Fisher School of Physics, Brandeis University, Waltham, MA 02453, United States of America}
\author{Richard Beals}
\email{richard.beals@yale.edu}
\affiliation{Department of Mathematics, Yale University, New Haven, CT 06520, United States of America}

\begin{abstract}
A variety of boundary value problems in linear transport theory are expressed as a diffusion equation of the two-way, or forward-backward, type. In such problems boundary data are specified only on part of the boundary, which introduces several technical challenges. Existence and uniqueness theorems have been established in the literature under various assumptions; however, calculating solutions in practice has proven difficult. Here we present one possible means of practical calculation. By formulating the problem in terms of projection operators, we derive a formal sum for the solution whose terms are readily calculated. We demonstrate the validity of this approach for a variety of physical problems, with focus on a periodic problem from the field of active matter.
\end{abstract}
\maketitle

\blfootnote{}
\blfootnote{\emph{This is an Accepted Manuscript. The journal reference for the final published version is J. Phys. A: Math. Theor. 52 115204. The Accepted Manuscript is available for reuse under a CC BY-NC-ND 3.0 licence after the 12 month embargo period provided that all the terms and conditions of the licence are adhered to.}}

\section{Introduction} \label{sec-introduction}

A number of physical systems are described by equations of the form
\be\label{basic}
h(\th) \pd{f(x,\th)}x\=-Af(x,\th),\quad 0 < x < L,
\ee
where $h$ has a change of sign and $-A$ is a non-negative self-adjoint differential or integral
operator in the $\th$ variable. These equations commonly describe particle transport in random media. One example is Bothe's equation for electron scattering
\be\label{bothe}
\sin \th \cos\th\,\pd{f(x,\th)}x\=\pd{}{\th}\left\{\sin\th\,\pd{f(x,\th)}\th\right\},
\ee
where $f(x,\theta)$ gives the distribution of electrons at position $x$ and with scattering angle $\theta$ \cite{Bothe1929}. Another widely studied example is the stationary Fokker--Planck equation
\be\label{FP}
v\,\pd{f(x,v)}x\=\gamma \frac{kT}m\,\pds{f(x,v)}v+\gamma \pd{}v\left[ vf(x,v) \right],
\ee
which describes the distribution of a Brownian particle in phase space \cite{Beals1983}. Here $v$ is the particle velocity, which plays the role of $\theta$ in equation \eqref{basic}: making the change $\theta \rightarrow v$, equation \eqref{FP} has $h(v) = v$ and
\begin{equation}
A = -\gamma \frac{kT}m\,\pds{}v-\gamma \pd{}v v.
\end{equation}
Other examples occur in such diverse subjects as astrophysics \cite{Freiling2003}, active matter\cite{Wagner2017}, neutron transport \cite{Case1960,McCormick1973}, and gas dynamics \cite{Cercignani1962,Cercignani1988}.

In our case we focus on problems in which $A$ is a differential operator of the Sturm--Liouville type:
\begin{equation}
Av(\th)\= -[p(\th)v'(\th)]'(\th).
\end{equation}
where $p(\theta)$ is a positive function. With this choice of $A$, equation \eqref{basic} has structure reminiscent of a diffusion equation, with $x$ playing the role of the time-like variable. On the other hand, the fact that the function $h(\theta)$ changes sign implies that ``forward diffusion" occurs where $h(\theta) > 0$ and ``backward diffusion" where $h(\theta) < 0$.  The structure of such equations is therefore fundamentally different from  that of ordinary diffusion-like equations, and leads to the terminology ``two-way'' or ``forward-backward'' problems. In particular, for the problem to be well-posed, we must specify initial conditions $f(0, \theta)$ only for values
of $\th$ where $h(\th)>0$, and final, or asymptotic, conditions $f(L,\theta)$ only where $h(\th)<0$. (Frequently, this can be interpreted as fixing particle fluxes incoming to the volume $x \in (0,L)$.) 

 The analysis of such problems typically begins by separating variables and 
considering the eigenvalue problem
\be\label{eigen1}
Au(\th)+\l h(\th)u(\th)\=0.
\ee
The success of this approach depends crucially on the completeness properties 
of the eigenfunctions. As we shall see in sections \ref{sec-theoreticalframework-I} and \ref{sec-theoreticalframework-II}, the nature of the initial
 and final conditions implies that the required property is that of 
{\it half-range completeness}, which asks 
whether the eigenfunctions with positive (resp.\ negative)
eigenvalues are complete when restricted to the ``half-range''
$\{\th:\,h(\th)>0\}$ (resp.\ $\{\th:\,h(\th)<0\})$.  The study of this 
half-range completeness question, as well as other aspects of the eigenvalue 
problem \eqref{eigen1}, is the subject
of an extensive literature (see Refs. \onlinecite{Mingarelli1986,Greenberg1987,Kaper1984,Beals1979,Beals1985,Beals1983} and
references therein). 
Besides their relevance for physical systems, these problems are interesting
because they generalize ``standard"  eigenvalue problems, such as those 
encountered in classical Sturm-Liouville theory, in which
the weight function $h(\theta)$ is assumed positive. 


Even when the half-range completeness property has been established,
a practical difficulty is that the half-range restrictions of the
eigenfunctions are not orthogonal with respect to any simply computed
inner product, precluding straightforward calculation of the coefficients in the expansion for $f(x,\theta)$. 
Progress in this direction has proven difficult; most previous work has focused on formal constructions of the half-range expansions
using complex variable techniques \cite{Marshall1985,Hagan1989,Hagan1999}. 
In this paper we present an alternative approach, based on a formulation of 
the problem in terms of projection operators, and leading
to a formal sum whose terms are easily calculated. In particular, the first few terms in the sum can often be given analytic expressions, leading to direct insights into the structure of the solution as a function of system parameters. Although proof of the convergence of the sum is not possible in all cases, numerical calculations establish conditions under which convergence holds; these conditions are satisfied by many physical systems.
This is illustrated with an application to the periodic problem
\bea
&&\cos\th\pd{f(x,\th)}x\= \pds{f(x,\th)}\th,\qquad 0<x<L;\label{example1}\\
&&f(0,\th)\=w(\th)\quad\hbox{where $\cos(\th)>0$};\label{example2}\\
&&f(L,\th)\=w(\th)\quad\hbox{where $\cos(\th)<0$}\label{example3},
\eea
which has been studied in the context of active matter \cite{Wagner2017}. 

\medskip

Many problems, including the periodic problem \eqref{example1} -- \eqref{example3}, have $0$ as an eigenvalue. On the other hand, the formalism is somewhat more complicated for the $0$ eigenvalue case than otherwise. Because of this, in section \ref{sec-theoreticalframework-I} we first develop the formalism for problems with only nonzero eigenvalues, before presenting in section \ref{sec-theoreticalframework-II} the $0$ eigenvalue case as an extension of the simpler formalism. (For details, and for a larger class of problems, see Ref. \onlinecite{Beals1979}). In section \ref{sec-convergence} questions of convergence are addressed. Finally, in section \ref{sec-formal-solution} we translate
our approach into a straightforward algorithm, which, in section \ref{sec-periodic-problem}, is applied 
to the problem \eqref{example1} -- \eqref{example3}. Readers interested primarily in application of the method may
find sections \ref{sec-formal-solution} and \ref{sec-periodic-problem} most relevant, though sections \ref{sec-theoreticalframework-I} and \ref{sec-theoreticalframework-II} provide some necessary background. 

For proofs of the assertions in sections \ref{sec-theoreticalframework-I} and \ref{sec-theoreticalframework-II}, see Refs. \onlinecite{Beals1979} and \onlinecite{Beals1985}. 

\section{Theoretical framework: simplest case} \label{sec-theoreticalframework-I} 

Throughout the paper, the general problem we are interested in is given by
\begin{equation}
Af(x,\th)+h(\th)f_x(x,\th)\=0,\quad a<\th<b,\ \ 0\le x\le L,\label{basic1}
\end{equation}
with boundary conditions specified by a function $w(\th)$: 
\bea
&&f(0,\th)\=w(\th)\quad\hbox{where $h(\th)>0$};\label{basic2}\\
&&f(L,\th)\= w(\th)\quad\hbox{where $h(\th)<0$}.\label{basic3}
\eea
Here  $h$ is a bounded function with finitely many zeros in the interval $a<\th<b$, and $A$ denotes the Sturm--Liouville operator 
\begin{equation}
Av(\th)\= -[p(\th)v'(\th)]'(\th),
\end{equation}
with $p(\th)>0$,
acting on functions $u$ that satisfy chosen boundary conditions, either
{\it separated\/}:
\be\label{separated}\cos{\a}\,v(a)+\sin\a\, v'(a)\=0\=
\cos{\b}\,v(b)+\sin\b\, v'(b),
\ee
or {\it  periodic\/}:
\be\label{periodic}
v(a)\=v(b),\quad v'(a)\=v'(b).
\ee
Let $\mathscr{H}$ be the space of continuous, piecewise twice differentiable 
functions that satisfy the 
chosen boundary conditions, and consider the eigenvalue problem
\eqref{eigen1}.  In this section we develop the simpler formalism corresponding to the case 
that $0$ is not an eigenvalue, before generalizing in section \ref{sec-theoreticalframework-II}. Thus, we number the eigenvalues
\be\label{numbering}
\cdots\le \l_{-3}\le\l_{-2}\le\l_{-1}<0<\l_1\le\l_2\le\l_3\le\cdots \ .
\ee
The eigenfunctions $u_j$,
\be\label{eigen}
Au_j(\th)+\l_jh(\th)u_j(\th)\=0, 
\ee
form a basis for $\mathscr{H}$.  Two integrations by parts show that
\bea\label{int1}
\l_j\int^b_a h(\th)u_j(\th)u_k(\th)\,d\th&=&-\int^b_a(Au_j(\th))u_k(\th)\,d\th\nonumber\\
&=&-\int^b_au_j(\th)Au_k(\th)\,d\th=\l_k\int^b_a u_j(\th)u_k(\th)h(\th)\,d\th.
\eea
Thus the $u_j$ are orthogonal with respect to the inner product
\be\label{ip1}
\la u,v\ra_A\=-\int^b_a (Au(\th))\,v(\th)\,d\th\=\int^b_a p(\th)u'(\th)v'(\th)\,d\th.
\ee
and we take them to be normalized: $\la u_j,u_j\ra_A=1$.  

In the case when $h$ changes sign, the $u_j$ are {\it not\/} orthogonal with respect to the 
natural inner product
\be
\la u,v\ra\=\lint u(\th)v(\th)|h(\th)|\,d\th.\label{ip2}
\ee
After rescaling, the $u_j$ lead to an orthonormal basis for an inner product
that is {\it equivalent\/} to $\la u,v\ra$.   Let
\begin{equation}
v_j(\th)\=|\l_j|^{-1/2}u_j(\th),
\end{equation}
so that
\begin{equation}
\la v_j,v_k\ra_A\=\sgn \l_j\,\d_{jk}\int^b_a v_j(\th)v_k(\th)h(\theta)d\th.
\end{equation}
Thus the  $v_j$ can be taken as an orthonormal basis for $\mathscr{H}$ with
respect to a new inner product $\la\,,\,\ra_1$: 
\be\label{ip1}
\la u,v\ra_1\=\int^b_a [P_+u(\th)-P_-u(\th)]\,v(\th)h(\theta)d\th.
\ee
where $P_\pm$ is the orthogonal projection onto the span of the $v_j$ for
which $\pm\l_j>0$, i.e. 
\bea
P_{\pm}v_j(\th)\=\left\{\begin{matrix} v_j(\th)&\  \hbox{if $\pm\l_j>0$}\\
0&\  \hbox{if $\mp\l_j<0$}\end{matrix}\right..
\eea

It is an important fact that the norms $||u||_1$ and $||u||$ defined
by the inner products $\la\,,\,\ra_1$ and $\la\,,\,\ra$ are equivalent:
there is a constant $C$ such that
\begin{equation}
C\inv ||u||\ \le\ ||u||_1\ \le\ C\,||u||,\quad\hbox{all $u\in \mathscr{H}$}.
\end{equation}

The {\it half--range completeness property\/} states that restrictions of the
eigenfunctions with eigenvalues $>0$ (resp.$<0$)  are a basis for functions
restricted to the range where $\pm h(\th)> 0$ (resp.$<0$) \cite{Beals1979,Beals1985}.  This, and 
the boundary conditions \eqref{basic2}, \eqref{basic3}, 
are most conveniently described using the projections $P_\pm$, which are
 orthogonal with respect to $\la\,,\,\ra_1$,
together with the projections $Q_\pm$, which are orthogonal with respect to $\la\,,\,\ra$:
\bea
Q_+u(\th)\=\left\{\begin{matrix} u(\th)&\  \hbox{if $h(\th)>0$}\\
0&\  \hbox{if $h(\th)<0$}\end{matrix}\right.;\qquad
Q_-u(\th)\=\left\{\begin{matrix}0&\  \hbox{if $h(\th)>0$}\\
u(\th)&\  \hbox{if $h(\th)<0$}\end{matrix}\right..
\eea
Half-range completeness means that $V(\mathscr{H})=\mathscr{H}$, where $V$ is the
operator
\begin{equation}
V\=Q_+P_++Q_-P_-.
\end{equation}
(Technical point: we have tacitly replaced the original $\mathscr{H}$ by its
completion with respect to the new inner product(s).)  The operator
$V$ is in fact an invertible map from $\mathscr{H}$ to $\mathscr{H}$ \cite{Beals1985}.

\smallskip
Now, suppose that $f$ is a solution of \eqref{basic1} -- \eqref{basic3}.  
Expanding
$f(x,\th)=\sum F_k(x)v_k(\th)$ for some functions $F_k$ leads to the form
\be\label{basic5}
f(x,\th)\=\sum_{\l_j>0}a_je^{-\l_jx}v_j(\th)+\sum_{\l_j<0}
a_je^{\l_j(L-x)}v_j(\th),
\ee
where the $a_j$ are independent of $x$. Next, define the function $v(\th)=\sum a_k v_k(\th)$.  The problem can then be
expressed in terms of $v$ by introducing operators $W$ and $M_L$:
\be\label{WML}
W\=Q_+P_-+Q_-P_+;\qquad M_Lv_j\=e^{-|\lambda_j| L}v_j.
\ee
Note that 
\begin{equation}
V+W\=(Q_++Q_-)(P_++P_-)\=(P_++P_-)\=I,
\end{equation}
where $I$ is the identity operator.  Then
\begin{equation} 
f(0,\th)\=P_+v(\th)+P_-M_Lv(\th),\quad f(L,\th)\=P_+M_Lv(\th)+
P_-v(\th), 
\end{equation}
so the boundary conditions can be written as $(V+WM_L)v=w$ or
\be \label{basic6}
(I+V\inv WM_L)v\=V\inv w.
\ee

It can be shown that the operator $V\inv WM_L$ has norm $<1$ with 
respect to the norm that
is associated to the inner product \eqref{ip1}:
\begin{equation}
||V\inv WM_L||_1\=\sup_{||u||_1=1}||V\inv WM_Lu||_1\ <\ 1.
\end{equation}
Therefore the solution to \eqref{basic1} -- \eqref{basic3} is given by
the Neumann series
\be\label{solution1}
v\=\left\{\sum^\infty_{n=0}(-V\inv WM_L)^n\right\} V\inv w.
\ee

For practical calculations, we must find a way to invert $V$. In principle this can be done approximately in $\mathscr{H}$ by restricting attention to the subspace spanned by the finite-dimensional set $\{v_j\}_{j=-N}^{j=N}$ for some $N$. Within this subspace, one can construct $V$ and invert it numerically on a computer. However, in many contexts it is desirable rather to generate a sequence of approximate analytic solutions to the problem. Such an approach is computationally less intensive and allows insight into the structure of the problem as a function of system parameters.

\smallskip
One approach along these lines is to reformulate the problem to get rid of the explicit operator inverse $V^{-1}$. We use the identity $V=I-W$ to rewrite equation \eqref{basic6} as
\begin{equation}
(V+WM_L)v\=(I-W_L)v\=w,\qquad W_L\=W-WM_L, \label{eq:sec-II-WL-definition}
\end{equation}
with formal solution
\be\label{alternate}
v\=\sum^\infty_{n=0}W_L^nw.
\ee
The question of convergence is discussed in section \ref{sec-convergence}. The advantage of this approach is that the first few terms in this series 
can often be computed by hand, generating a sequence of approximate 
analytic solutions to the problem.

\section{Theoretical framework,  extended}\label{sec-theoreticalframework-II}

In preparation for treating the illustrative example \eqref{example1} --
\eqref{example3}, we modify the assumptions in section \ref{sec-theoreticalframework-I} in two ways.
First, we assume that $h$ has mean value zero:
\be\label{mean0}
\lint h(\th)\,d\th \=0.
\ee
This corresponds to a symmetry between forward and backward scattering, and is satisfied by many physical systems, including those mentioned in the introduction. 

Second, we assume that $0$ is an eigenvalue, i.e.\ that the constant 
function $1$  satisfies the boundary conditions.  (This is true in
the periodic case; it is also true in the separated
case for Neumann boundary conditions: $u'(a)=u'(b)=0$.) Eigenvectors $u_j$ corresponding to non-zero eigenvalues are chosen
as before. Under these assumptions the constant function and the $u_j$ are actually not complete in $\mathscr{H}$. In addition to these we need the function $g$, the unique solution to 
\be\label{g}
Ag(\th)\=h(\th),\qquad\int^b_a g(\th)\,d\th\=0.
\ee
It will prove convenient to treat $g$ and the constant function $1$
separately from the eigenfunctions $v_j$ discussed above. 
For this purpose we take $\mathscr{H}_1$ to be the subspace of $\mathscr{H}$ consisting of functions
$u$ such that
\be\label{1g}
\int^b_a u(\th)h(\th)\,d\th\=0\=\int^b_a u(\th)g(\th)h(\th)\,d\th.
\ee
The $u_j$ (which satisfy these constraints, as is easily checked)
are an orthogonal basis for $\mathscr{H}_1$ with respect to the inner product \eqref{ip1}.  By renormalizing
to $\{v_j\}$ as before, we obtain an orthonormal basis for $\mathscr{H}_1$
with respect to the inner product \eqref{ip1}.

\smallskip
Expanding a proposed solution of \eqref{basic1}, taking into account
the additional functions $1$, $g$ needed to span $\mathscr{H}$, we find
two new solutions
\begin{equation}
f_1(x,\th)\ = 1;\qquad  f_2(x,\th)\=x + g(\th).
\end{equation}
Thus the full expansion is
\be\label{basic7}
f(x,\th)\=c+d(x+g(\th))+\sum_{\l_j>0}a_je^{-\l_jx}v_j(\th)+\sum_{\l_j<0}
a_je^{\l_j(L-x)}v_j(\th).
\ee
This leads to a corresponding modification of the results in the previous
section: the boundary conditions $w$ for $f$ lead to
\be\label{neweqn}
w\=c+d g_L+(V+WM_L)v\=c+dg_L+(I-W_L)v,\qquad v\in \mathscr{H}_1.
\ee
Here
\begin{equation}
g_L(\th)\=\left\{\begin{matrix} g(\th)&\ \ {\rm if}\ \ h(\th)>0;\\
L + g(\th)&\ \ {\rm if}\ \ h(\th)<0.\end{matrix}\right.
\end{equation}
As before $V=Q_+P_++Q_-P_-$ and $W=Q_-P_++Q_+P_-$, where $P_\pm:\mathscr{H}_1\to \mathscr{H}_1$.

Let $\mathscr{H}_0$ denote the subspace spanned by $1$ and $g_L$, and let $P$ be the
projection of $\mathscr{H}$ to $\mathscr{H}_1$ that vanishes on $\mathscr{H}_0$.   Applying $P$ to
\eqref{neweqn} gives an equation in the subspace $\mathscr{H}_1$:
\be\label{neweqn2}
(I-PW_L)v\=Pw, 
\ee
with formal solution
\begin{equation} 
v=\sum_{n=0}^{\infty} (PW_L)^n Pw.
\label{sec3-formal}
\end{equation} 
\section{Convergence of the formal solution} \label{sec-convergence}

\begin{figure}
\centering
  \includegraphics[width=81.5mm,height=52.5mm]{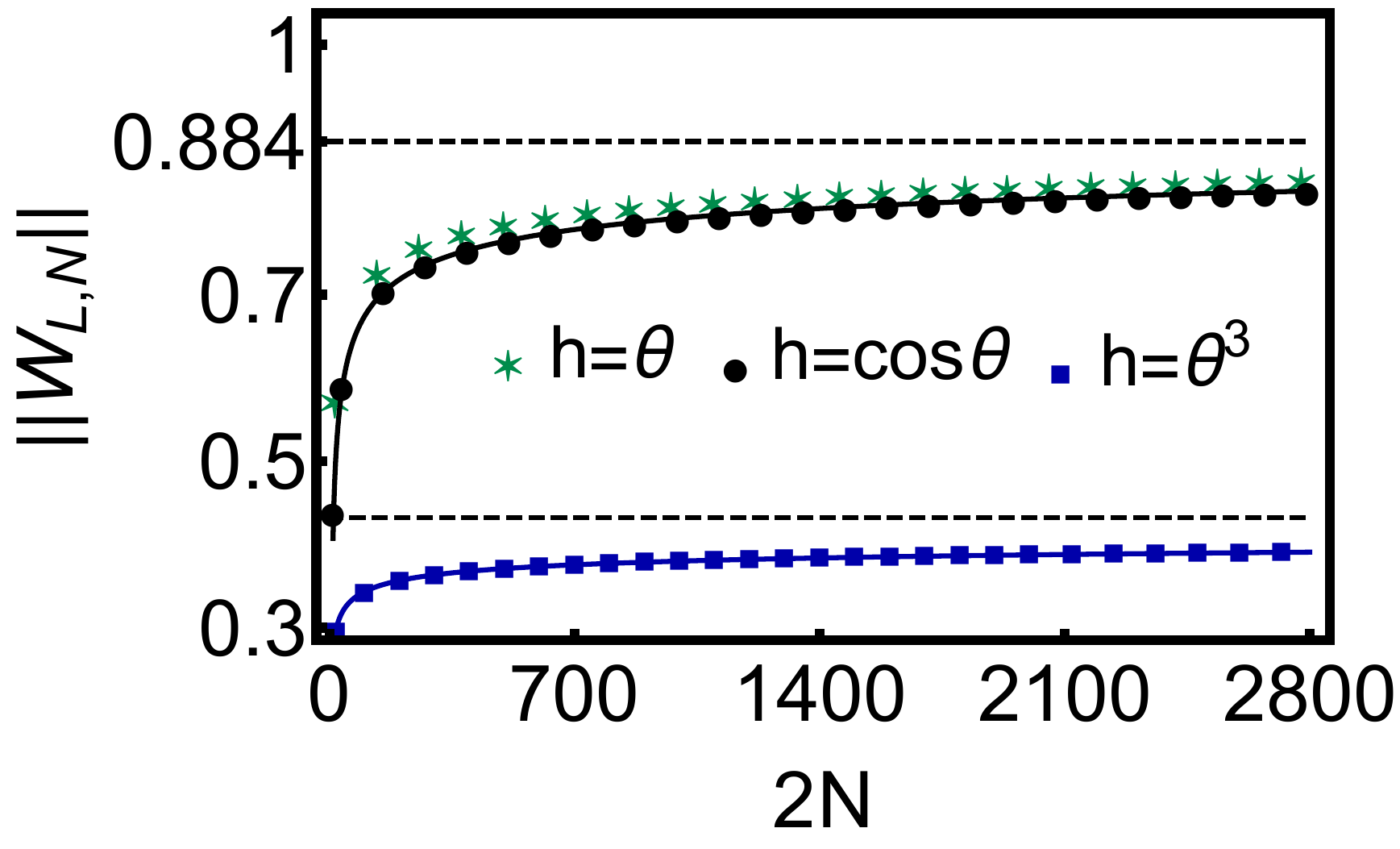}  \hspace{5mm}
    \includegraphics[width=75mm,height=52.5mm]{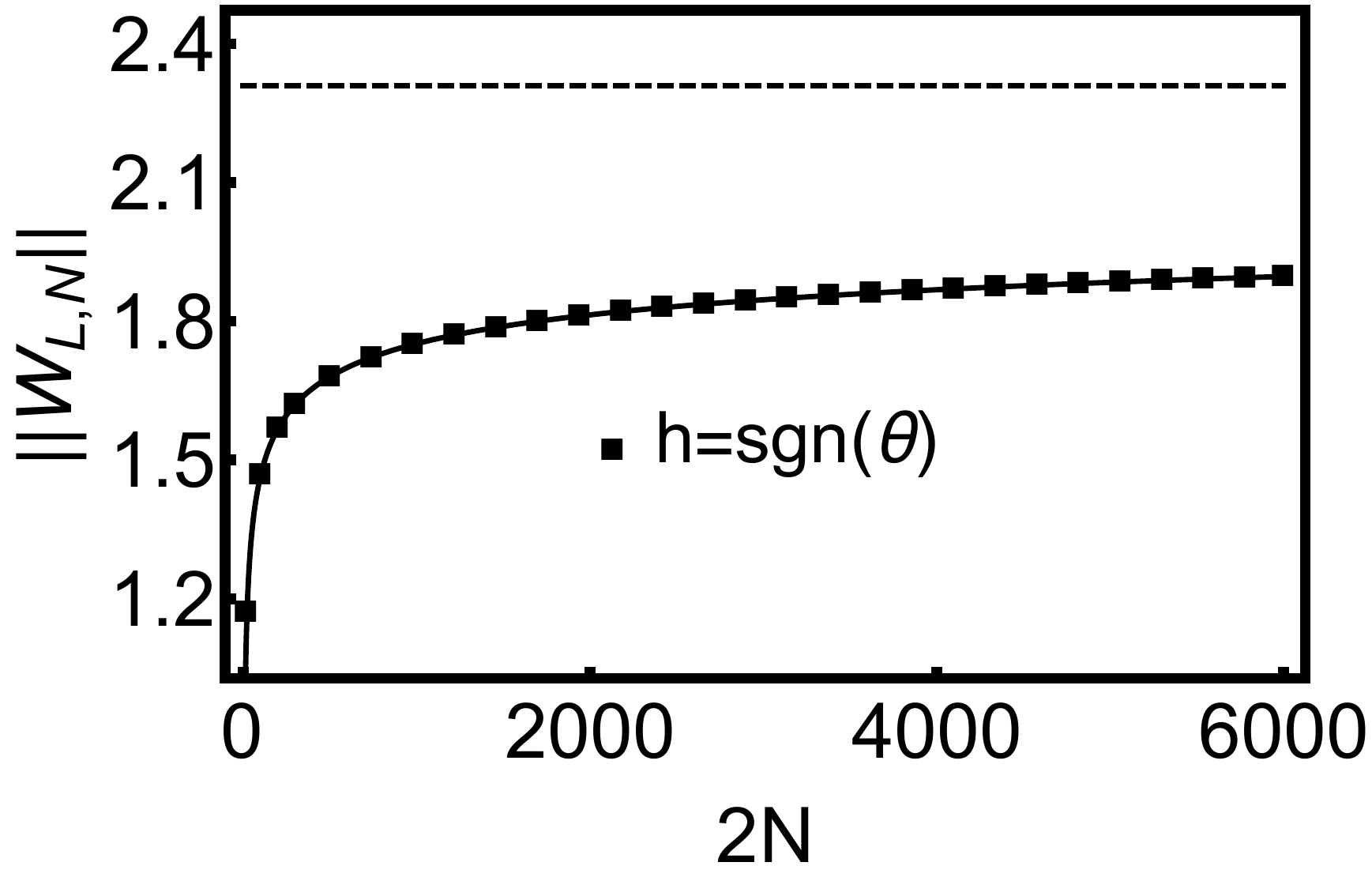}
  \caption{Operator norm $||W_{L, N}||$ as a function of the number of eigenfunctions $2N$ and for various choices of $h(\theta)$.  For convenience, the transcendentally small $L$ dependence is dropped: inclusion of this contribution only modifies $||W_{L, N}||$ downwards (see appendix B). The $h(\theta) = \cos \theta$ case uses $2 \pi$-periodic boundary conditions, and all others use absorbing boundary conditions on $-1 < \theta < 1$. Power law fits are shown in solid black and blue, together with the corresponding asymptotes as $N \rightarrow \infty$.}
  \label{fig:w_estimates}
\end{figure}

Here the relevant criterion for the convergence of \eqref{sec3-formal} is whether the norm of $PW_L$ 
with respect to the $L^2$ inner product $\la\ ,\ \ra$ is less than 1:
\begin{equation}
||PW_L|| < 1. \label{eqn:convergence-criterion}
\end{equation} 
Using $||PW_L|| \leq ||P||\cdot||W_L||$, a sufficient criterion is
\begin{equation}
||P||\cdot||W_L|| < 1.
\end{equation}
The quantity $||P||$ is straightforward to calculate. For instance, in appendix C it is shown that $||P|| = (4 \sqrt{6}) / (3 \pi) = 1.0395...$ for the periodic problem \eqref{example1} -- \eqref{example3}. By contrast, the norm on $W_L$ is harder to estimate. With a WKB-type analysis it is possible to get an upper bound in terms of the eigenvalue spectrum (appendix A); however, in many cases this bound is insufficient to ensure equation \eqref{eqn:convergence-criterion}.

In practice, one is interested anyway in a slightly different calculation, namely approximation of each term in \eqref{sec3-formal} using a finite number of eigenfunctions. In this case one can demonstrate convergence of the sum numerically. More precisely, we approximate the sum \eqref{sec3-formal} by restricting attention to the space $\mathscr{H}_{1,\leq N}=\mathrm{span}\{v_{j}\}_{\left\vert j\right\vert \leq N}$, i.e. the span of the first $2 N$ eigenfunctions ordered by the magnitude of their eigenvalues. The series \eqref{alternate} (simple framework, section \ref{sec-theoreticalframework-I}) then becomes
\begin{equation}
v\left( \theta \right) =\sum_{n=0}^{\infty }\left(P_{N}W_{L, N}\right)
^{n}P_Nu\left( \theta \right), \label{finite-N-formal-sol}
\end{equation}
where $W_{L, N}:\mathscr{H}_{1,\leq N} \to \mathscr{H}_1$ is the finite-dimensional analogue of $W_L$ and satisfies
\begin{equation}
W_{L, N}(\mathscr{H}_{1,\leq N}) = W_L (\mathscr{H}_{1,\leq N});
\end{equation}
and $P_{N}$ is the projection operator defined by
\begin{equation}
P_{N}(\mathscr{H}_{1,\leq N})=\mathscr{H}_{1,\leq N} \qquad \text{and} \qquad P_{N}\left(\mathscr{H}_{1,>N}\right) =0, 
\end{equation}
with $\mathscr{H}_{1,>N} =  \mathrm{span}\{v_{j}\}_{\left\vert j\right\vert >N}$. The corresponding expression for the extended framework can similarly be obtained from equation \eqref{sec3-formal}.


As shown in appendix B, $||W_{L, N}||$ can be evaluated numerically for not too large $N$. The results for various $h(\theta)$ are shown in figure \ref{fig:w_estimates} for $A = -\partial^2 / \partial \theta^2$. In particular, we note that for the periodic problem \eqref{example1} -- \eqref{example3}, $||P|| \cdot ||W_{L, N}|| < 1$ for values of $N$ used in most practical calculations. More generally, the scaling of $||W_{L, N}||$ appears predominantly determined by the algebraic multiplicity of $h(\theta)$ at the turning point. This is confirmed by the WKB-type analysis in appendix A, and suggests that any problem with a linear or cubic turning point and $\int h(\th)\,d\th \=0$ has $||P|| \cdot ||W_{L, N}|| < 1$ for reasonably sized $N$. Briefly, we note also that the numerical data fit well to a power law. Extrapolating to large $N$ suggests in fact that $||W_L||$ (not just $||W_{L, N}||$) is less than $1$ for linear and cubic turning points. For practical calculations, however, the estimates on $||W_{L, N}||$ are sufficient.

Finally, to complete the argument we need an estimate on $P_N$. Here we simply point out that on physical grounds, solutions to problems like \eqref{example1} -- \eqref{example3} are expected to be reasonably smooth and slowly varying. In this case modes with large eigenvalues do not contribute significantly, and $||P_N|| \approx 1$. (We note that suppression of higher order modes is required anyway to justify the use of the first $2 N$ eigenfunctions for a given $N$.) 

\section{Alternate formulation for eigenvalues with small magnitude} \label{sec-alternate-formulation}

The presence of small, nonzero eigenvalues may cause the formal solution \eqref{sec3-formal} to diverge. To explain why, and to motivate an alternate formulation which may restore convergence, we consider the illustrative problem
\bea
&&(\cos\th - r)\pd{f(x,\th)}x\= \pds{f(x,\th)}\th,\qquad 0<x<L; \label{sec5-periodic-generalization-1}\\
&&f(0,\th)\=w(\th)\quad\hbox{where $\cos(\th) - r>0$};\label{sec5-periodic-generalization-2}\\
&&f(L,\th)\=w(\th)\quad\hbox{where $\cos(\th) - r<0$},\label{sec5-periodic-generalization-3}
\eea
where $0 \leq r < 1$. This is a generalization of the periodic problem \eqref{example1} -- \eqref{example3} with applications in the field of active matter. It is solved in detail in Ref. \onlinecite{Wagner2017}.

Numerically, one finds that the norm of $PW_L$ as defined in section \ref{sec-theoreticalframework-II} diverges as $r \rightarrow 0$. The reason can be traced to the spectral structure of the problem for $r \neq 0$: besides $0$, there is an eigenvalue $\lambda_R = 2 r + \mathcal{O}(r^3)$ which merges with the $0$ eigenvalue as $r \rightarrow 0$. All other eigenvalues have positive magnitude in the limit $r \rightarrow 0$. The eigenvalue $\lambda_R$ in particular causes issues in two ways. 

\begin{figure}
\centering
  \includegraphics[width=81.5mm,height=52.5mm]{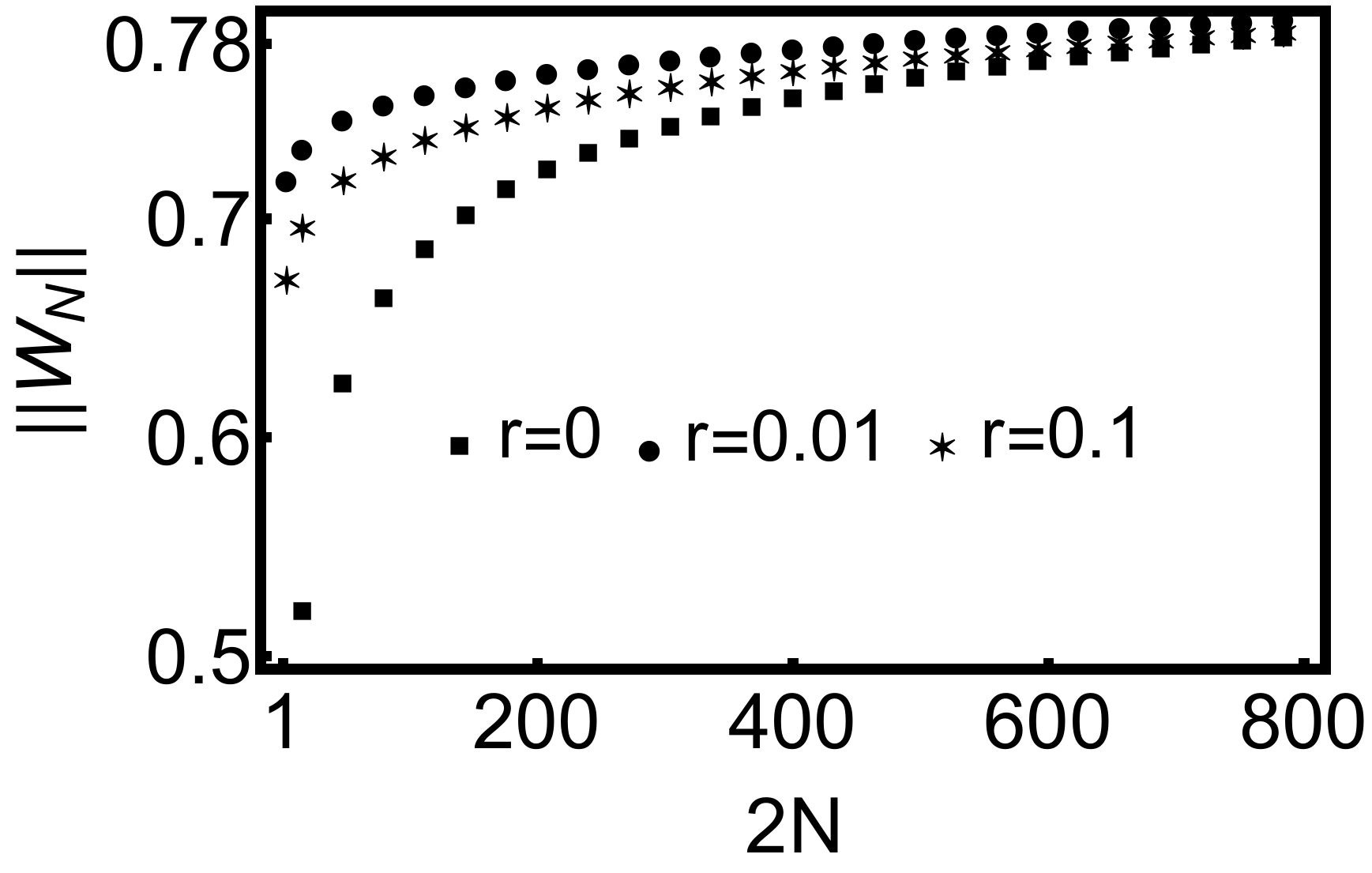}  \hspace{5mm}
    \includegraphics[width=75mm,height=52.5mm]{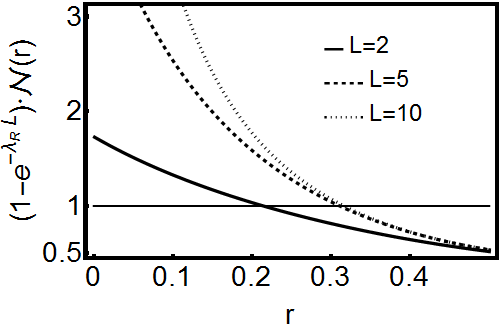}
  \caption{Estimates related to the problem \eqref{sec5-periodic-generalization-1} - \eqref{sec5-periodic-generalization-3}. The plot on the left shows the modest increase in $||W_N||$ for small but nonzero $r$. The plot on the right shows the lower bound on $||W_L P||$ from equation \eqref{eq:wr-lower-bound} for different values of $L$. Values of $r$ and $L$ for which this bound is greater than $1$ indicate the divergence of the formal solution \eqref{sec3-formal}.}
  \label{fig:wr_norm_estimates}
\end{figure}

First, the WKB analysis in appendix A shows that estimates on the norm of $W$ tend to grow larger as the magnitude of the nonzero eigenvalues decreases. For the problem \eqref{sec5-periodic-generalization-1} -- \eqref{sec5-periodic-generalization-3}, the fact that $\lambda_R \rightarrow 0$ as $r \rightarrow 0$ is expected to correspond to an increase in $||W||$ or $||W_N||$ (note that this is $W$ rather than $W_L$; see equation \eqref{eq:sec-II-WL-definition}). Although not dramatic, this increase is indeed observed (figure \ref{fig:wr_norm_estimates}). 

Second, the merging of $\lambda_R$ with the $0$ eigenvalue causes $||P||$ to diverge. This is because the eigenfunction corresponding to $\lambda_R$ (call it $v_R$) has the form $v_R = 1 + 2 r \cos \theta + \mathcal{O}(r^2)$. Recalling that the eigenfunction corresponding to the $0$ eigenvalue is just $1$, we can then write
\begin{equation}
||P|| \geq \frac{||P(v_R - 1)||}{||v_R - 1||} = \frac{||1 + 2 r \cos \theta + \mathcal{O}(r^2)||}{||2 r \cos \theta + \mathcal{O}(r^2)||} \sim \frac{1}{r}.
\end{equation}


Motivated by these observations, we write the following estimate on the combined product $W_L P$:
\begin{equation}
||W_L P|| \geq \frac{||W_L P(v_R - 1)||}{||v_R - 1||} = (1 - e^{-\lambda_R L}) \cdot \mathcal{N}(r), \label{eq:wr-lower-bound}
\end{equation}
where 
\begin{equation}
\mathcal{N}(r) = \sqrt{\frac{\int_{\cos \theta  < r} (v_R)^2 |\cos \theta - r| d\theta}{\int_{-\pi}^{\pi}  (v_R - 1)^2 |\cos \theta - r| d\theta}}.
\end{equation}
As $r \rightarrow 0$, $\mathcal{N}(r) \sim 1 / r$ and  $ (1 - e^{-\lambda_R L}) \sim r L$, so that the bound on $||W_L P||$ approaches an $L$-dependent value. Numerical results for various $L$ are shown in figure \ref{fig:wr_norm_estimates}. For moderately sized $L$ and sufficiently small $r$, $||W_L P||$ is greater than $1$. 

The problematic nature of $\lambda_R$, and in general of any eigenvalue with small magnitude, suggests the treatment of any small-eigenvalue eigenfunctions separately in their own subspace rather than acting on them with $W_L$. More precisely, define
\begin{equation}
\overline{v}_{j}(\th)=\left\{
\begin{array}{cc}
v_{j}(\th), & \text{if}\ \sgn(\lambda _{j})h(\theta )>0; \\
e^{-|\lambda _{j}|L}v_{j}(\theta), & \text{if}\ \sgn(\lambda _{j})h(\theta )<0.
\end{array}                                                                 
\right.
\end{equation}
Then \eqref{neweqn} can be written
\be\label{neweqn3}
c+d g_L+\sum^{\infty}_{-\infty}a_j\overline{v_j}\=w.
\ee
Given $\bl>0$, let 
\begin{equation}
\mathscr{H}_{0,\bl}\={\rm span}\{1,g_L,{\overline v}_{j}\}_{|\l_j| < \bl};\qquad
\mathscr{H}_{1,\bl}\={\rm span}\{{\overline v}_{j}\}_{|\l_j| \ge \bl}.
\end{equation} 
We take $P_{\bl}$ to be the projection onto $\mathscr{H}_{1,\bl}$ that vanishes 
on $\mathscr{H}_{0,\bl}$.  Applying $P_\bl$ to \eqref{neweqn3} gives
\be\label{neweqn4}
(I-P_{\bl}W_L)v_\bl\=P_\bl w,\qquad v_\bl\=\sum_{|\l_j|\ge\bl}a_j v_j
\in \mathscr{H}_{1,\bl},
\ee
which is the analogue of equation \eqref{neweqn2}. While we do not attempt to analyze the convergence of this alternate scheme for general problems, we do indeed observe that it restores convergence for the problem \eqref{sec5-periodic-generalization-1} -- \eqref{sec5-periodic-generalization-3}. 


%
%
%

\section{Calculation of the Formal Solution} \label{sec-formal-solution}

In this section we outline the steps for computing terms in the formal solution \eqref{sec3-formal}:
\begin{equation}
v=\sum_{n=0}^{\infty} (PW_L)^n(Pw).
\label{formal-sec4}
\end{equation}
The first term $Pw$ can be calculated by expanding:
\begin{equation}
w = c_0 + d_0 g_L +  \sum_{\l_j>0}a^0_jv_j(\th)+\sum_{\l_j<0}a^0_jv_j(\th).
\label{section4-eigenbasis-expansion}
\end{equation}
Here $c_0$ and $d_0$ are determined from
\begin{equation}
\int \left(w - c_0 - d_0 g_L\right) h d\theta = 0 = \int \left(w - c_0 - d_0 g_L\right)ghd\theta,
\label{sec4-coeff1}
\end{equation}
whereas
\begin{equation}
a^0_j = \text{sgn}(\lambda_j) \int \left(w - c_0 - d_0 g_L\right) v_j h d\theta.
\label{sec4-coeff2}
\end{equation}
Comparing with \eqref{basic7} and \eqref{neweqn}, the zeroth order solution is then
\begin{equation}
f_0(x,\th)\=c_0+d_0\,(x + g(\th))+\sum_{\l_j>0}a^0_je^{-\l_jx}v_j(\th)+\sum_{\l_j<0}a^0_j
e^{(L-x)\l_j}v_j(\th).
\end{equation}
We note that the expressions for $c_0$ and $d_0$ are the same as those obtained by Bethe, et al in the context of electron scattering \cite{Bethe1938}. There the motivation is physical: since the average flux of incident electrons is more physically relevant than their exact distribution, Bethe et al choose to satisfy the boundary conditions only in the mean. By comparison, here we have re-derived this approximation as the first term in a systematic expansion.

Moving beyond the first approximation, the next term is $PW_LPw$. We already know 
\begin{equation}
W_LPw=\left\{
\begin{array}{cc}
\sum_{\l_j<0}a_j^0(1 - e^{\lambda _{j} L})v_j  & h(\theta) >0; \\
\sum_{\l_j>0}a_j^0(1 - e^{-\lambda _{j} L})v_j  & h(\theta) <0. \label{sec4-error-term}
\end{array}%
\right.
\end{equation}
The quantity $PW_LPw$ can be computed by expanding:
\begin{equation}
W_LPw = c_1 + d_1 g_L +  \sum_{\l_j>0}a^1_jv_j(\th)+\sum_{\l_j<0}a^1_jv_j(\th),
\label{section4-eigenbasis-expansion1}
\end{equation}
the coefficients determined from equations \eqref{sec4-coeff1} and \eqref{sec4-coeff2}, with $w$ replaced by $W_LPw$. The first order solution is then
\begin{align}
f_1(x,\th)&=(c_0+c_1)+(d_0+d_1)\,(x + g(\th))+\sum_{\l_j>0}(a^0_j+a^1_j)e^{-\l_jx}v_j(\th) \nonumber \\
&+\sum_{\l_j<0}(a^0_j+a^1_j)
e^{(L-x)\l_j}v_j(\th).
\end{align}
We can now proceed systematically. At the $n^{th}$ step of the procedure, the results from the previous step are used to expand $(W_LP)^nw$ in the same form as \eqref{section4-eigenbasis-expansion1}. The $n^{th}$ order coefficients $c_n$, $d_n$, and $a_j^n$ can be read off. Then, the coefficients in equation \eqref{basic7} are given by $c = \sum_n c_n$, $d = \sum_n d_n$, and $a_j = \sum_n a^n_j$.


\section{Application to a periodic problem} \label{sec-periodic-problem}
We illustrate these techniques in the context of the problem
\bea
&&\cos\th\pd{f(x,\th)}x\= \pds{f(x,\th)}\th,\qquad 0<x<L;\label{sec5-example1}\\
&&f(0,\th)\=w(\th)\quad\hbox{where $\cos(\th)>0$};\label{sec5-example2}\\
&&f(L,\th)\=w(\th)\quad\hbox{where $\cos(\th)<0$}\label{sec5-example3}.
\eea

\begin{figure}
\centering
  \includegraphics[width=0.5\linewidth,height=0.3734939760\linewidth]{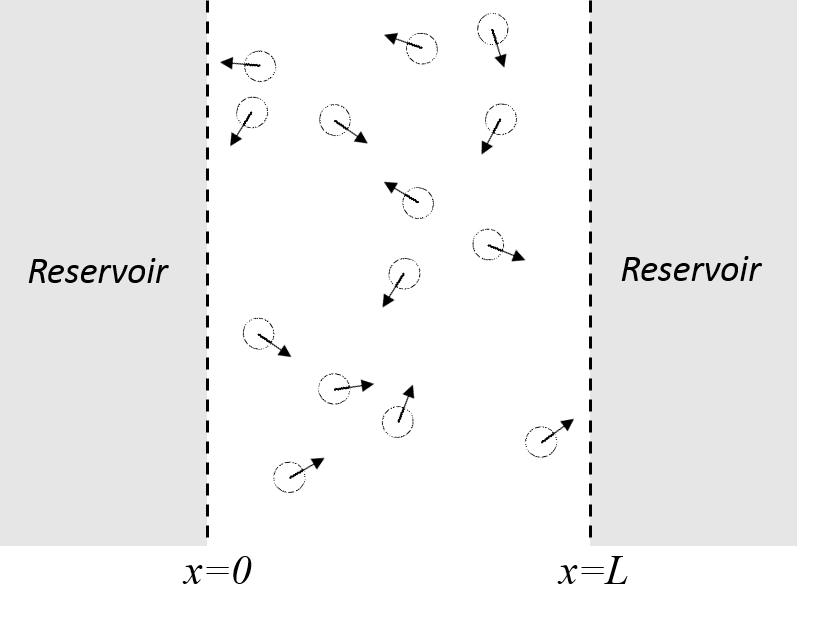}
  \caption{The channel geometry for active Brownian particles. The system is infinite in the vertical direction and of uniform width $L$ in the horizontal ($x$) direction. Reservoirs of particles are situated at $x=0$ and $x=L$.}
  \label{fig:infinite_channel}
\end{figure}

This equation occurs in the study of active matter, where it gives the steady-state distribution of active Brownian particles (ABPs) in certain 2d geometries \cite{Wagner2017}. In the ABP model, particles self-propel with constant velocity along an internal axis parametrized by angle $\theta \in (0, 2 \pi )$. The internal axis in turn reorients itself via diffusion in the $\theta$ variable. The resulting trajectories are correlated over a characteristic length, which in the non-dimensionalized model considered here is equal to $1$. The equation \eqref{sec5-example1} gives the steady-state distribution $f(x,\theta)$ of ABPs in cases where symmetry reduces the problem to an effective 1d description, e.g. particles in an infinite channel (figure \ref{fig:infinite_channel}). In such cases the self-propulsion is represented by the streaming term $\cos \theta \left( \partial /\partial x\right) $, whereas the orientational diffusion corresponds to $\partial^2 / \partial \theta^2$. Problems of this type are also discussed in Ref. \onlinecite{Wagner2017}. 

We consider the case $w(\theta) = \rho_1$ where $\cos \theta > 0$ and $w(\theta) = \rho_2$ where $\cos \theta < 0$. Here $\rho_1$ and $\rho_2$ are independent of $\theta$ and represent uniform reservoirs of ABPs: the one on the l.h.s. of the channel having density $2 \pi \rho_1$, and the one on the r.h.s. having density $2 \pi \rho_2$. Our starting point is the expansion
\be\label{expansion}
f(x,\th)\=c+d\,(x - \cos \th)+\sum_{\l_j>0}a_je^{-\l_jx}v_j(\th)+\sum_{\l_j<0}a_j
e^{(L-x)\l_j}v_j(\th).
\ee
One interesting physical quantity is the net particle flux in the channel due to different densities in the left and right reservoirs. Owing to the orthogonality properties of the $v_j$, this flux is independent of $x$ and directly proportional to $d$:
\begin{align}
\text{flux} &= \int f(x,\theta) \cos \theta d\theta \\
&= -\pi d .\label{flux_expression}
\end{align} 
Thus, obtaining an analytical approximation for $d$ will give insight into the transport behavior of ABPs. 

Using the results in section \ref{sec-convergence}, it is clear that the formal solution \eqref{sec3-formal} converges using any realistic number of eigenfunctions. In fact, the numerical results suggest that convergence is obtained on the full functional space as well, though this fact is not needed for practical calculation. Here we calculate $c$ and the $a_j$ through $n=1$ in equation \eqref{sec3-formal} and $d$ through $n=2$. Defining $\Delta \rho = \rho_2 - \rho_1$, the results for $c$ and $d$ are summarized as follows:
\begin{align}
c &\simeq \frac{\rho_1 + \rho_2}{2} -  \left[\frac{1 - \mathcal{A}(L)}{2} \right] \left(\frac{2L}{2L + \pi}\right) \Delta \rho - \frac{\mathcal{A}(L)}{2} \left(\frac{2L}{2L + \pi}\right)^2 \Delta \rho; \\
d \cdot \frac{L}{\Delta \rho} &\simeq \left[1 - \mathcal{A}(L) + \mathcal{B}(L) \right] \left(\frac{2L}{2L + \pi}\right) + \left[ \mathcal{A}(L)^2 + \mathcal{A}(L) - \mathcal{B}(L) \right] \left(\frac{2L}{2L + \pi}\right)^2 \label{d_arb_L} \nonumber \\
&- \mathcal{A}(L)^2 \left(\frac{2L}{2L + \pi}\right)^3 .
\end{align}

Details of the calculation as well as the complete expressions for $\mathcal{A}(L)$, $\mathcal{B}(L)$, and the $a_j$ are given in appendix C. The $L$ dependence of the quantities $\mathcal{A}(L)$ and $\mathcal{B}(L)$ is not easy to discern for arbitrary $L$. However, the expressions simplify considerably if we assume $L \gg 1/\lambda_1 \approx 0.094$, in which case $\mathcal{A}(L)$ and $\mathcal{B}(L)$ rapidly approach constant values:
\begin{align}
\mathcal{A}(L) \longrightarrow \mathcal{A}_0 \approx 0.070; \\
\mathcal{B}(L) \longrightarrow \mathcal{B}_0 \approx 0.035.
\end{align}
Then we find
\begin{align}
c &\simeq \frac{\rho_1 + \rho_2}{2} -  0.47 \left(\frac{2L}{2L + \pi}\right) \Delta \rho - 0.035 \left(\frac{2L}{2L + \pi}\right)^2 \Delta \rho \label{c_large_L}, \\
d \cdot \frac{L}{\Delta \rho} &\simeq 0.97 \left(\frac{2L}{2L + \pi}\right) + 0.04 \left(\frac{2L}{2L + \pi}\right)^2 -0.005 \left(\frac{2L}{2L + \pi}\right)^3. \label{d_large_L}
\end{align}

\begin{figure}
\centering
\subfloat[Density $g(x) = \int f(x,\theta) d\theta$.]{
\includegraphics[width=70mm,height=52.5mm]{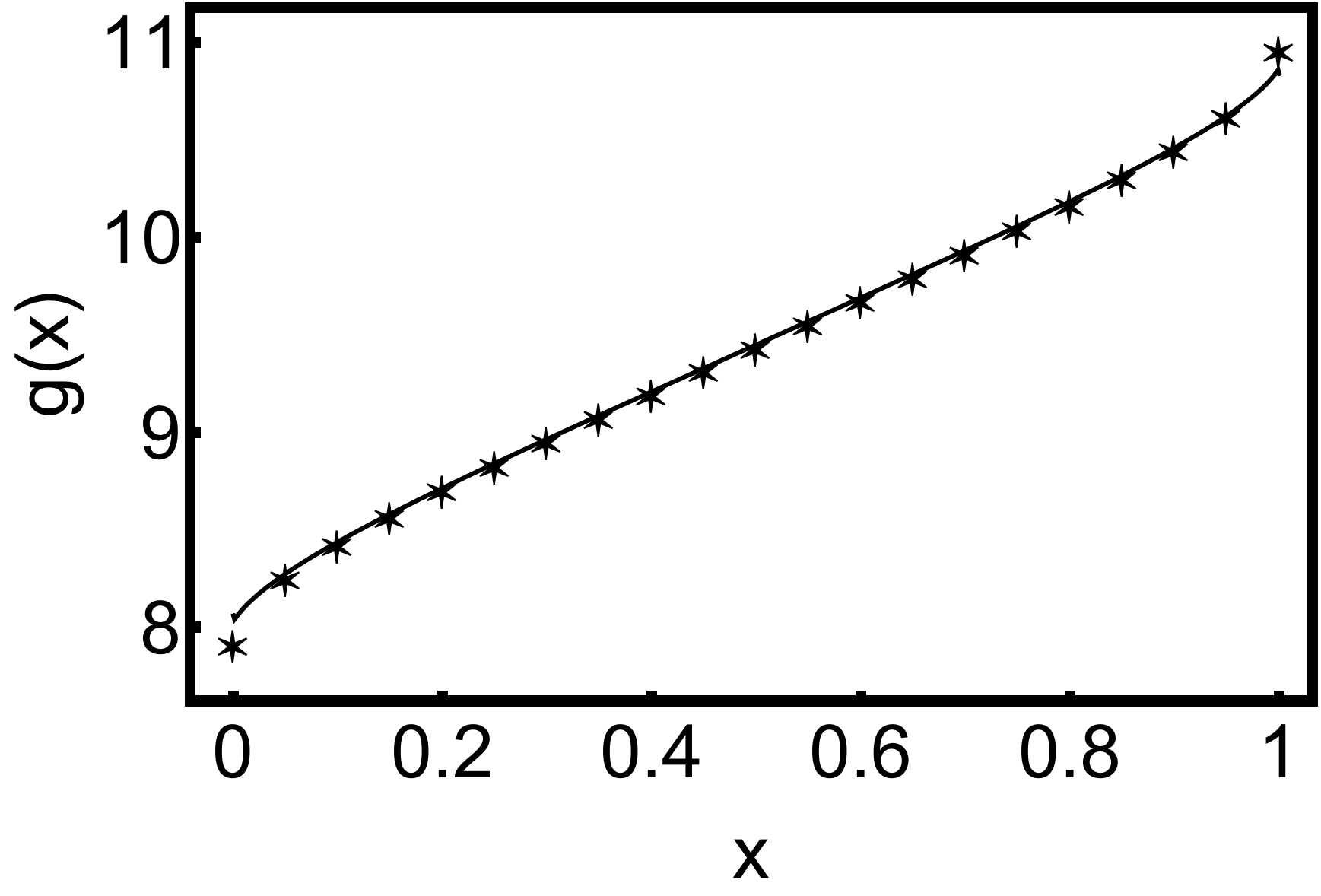} \hspace{10mm}}
\subfloat[The orientational distribution of particles exiting the channel at $x = L$. In terms of $f$, this is $Z^{-1} f(L,\theta)$, where $Z$ is chosen so the distribution is normalized to 1 on the range shown.] {
\includegraphics[width=78mm,height=52.5mm]{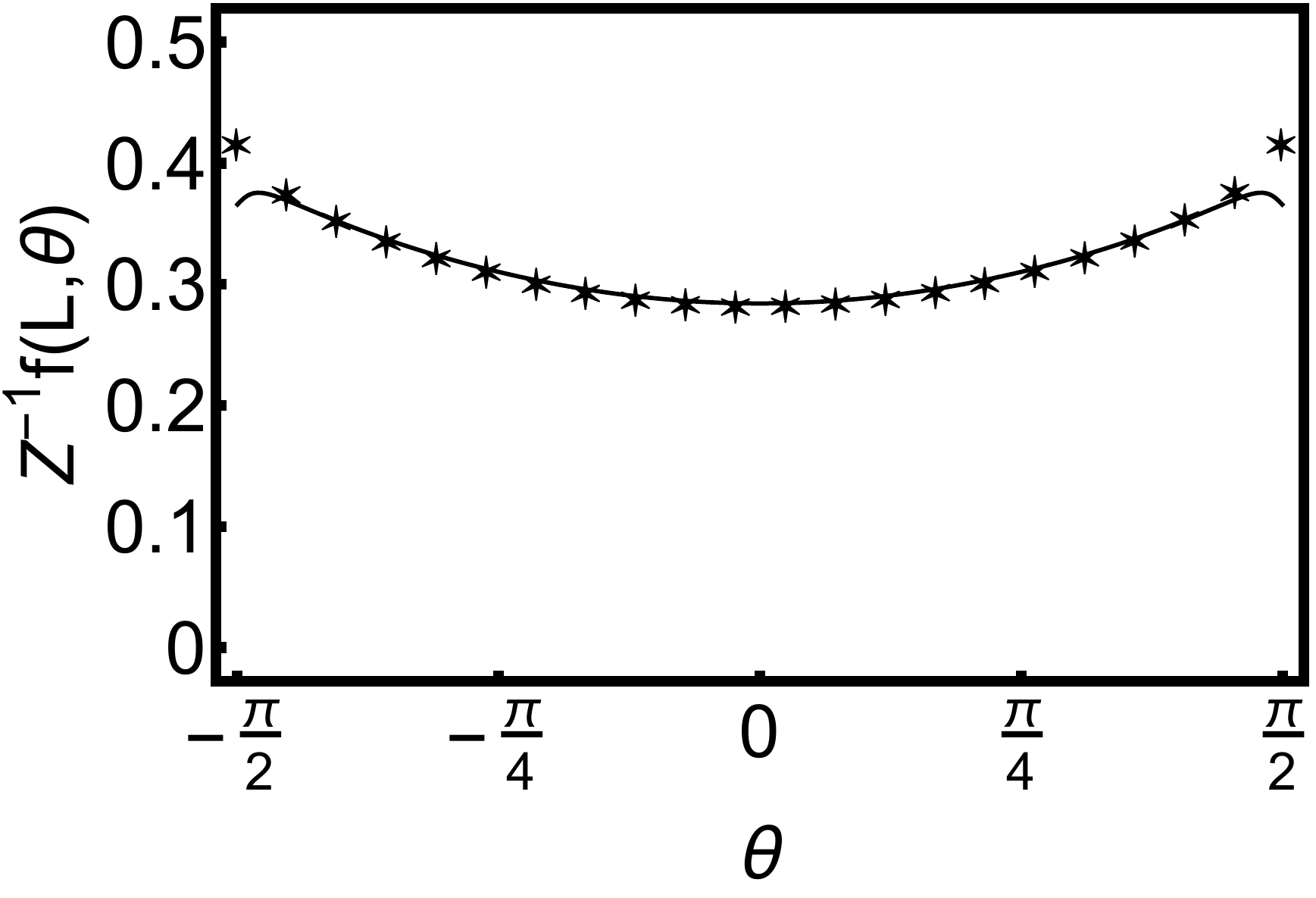}}
\caption{Comparison of the exact solution (asterisks) with the approximate solution (solid black) derived in the main text. Here $\rho_2 = 2$, $\rho_1 = 1$, and $L=1$. The exact solution is obtained by evaluating the formal sum \eqref{sec3-formal} numerically to arbitrary precision. The black curve shows the approximate solution determined by equations \eqref{c_large_L} and \eqref{d_large_L} for $c$ and $d$ and the expression for $a_j$ given in appendix C.}
\label{fig:solution-compare}
\end{figure}


Calculating to higher order in equation \eqref{sec3-formal} will continue to generate corrections in powers of $2L/(2L + \pi)$. However, already the third order correction to $d$ is small; hence, the above approximations are sufficient for most purposes. Indeed, as shown in figure \ref{fig:solution-compare}, comparison with the exact solution shows good agreement.

More importantly, the analytical expressions for the expansion coefficients allow insight into the physics of the problem. Here we focus on the coefficient $d$, which is proportional to the net flux in the channel via equation \eqref{flux_expression}. Examination of equations \eqref{d_arb_L} and \eqref{d_large_L} reveals several qualitatively distinct regimes, which shed light on the transport behavior of ABPs. 


\vspace{5mm}

$\mathbf{L \ll 1/\lambda_1 \approx 0.094}$ 

\vspace{2 mm}

For very small $L$, $\mathcal{A}(L)$ and $\mathcal{B}(L)$ both go to zero, and $d/\Delta \rho$ becomes a constant. In this case, particles travel directly from one reservoir to another without any chance for their axis of self-propulsion to reorient. Thus, the transport in the channel is ballistic, and the flux does not depend on the width $L$.

\vspace{2 mm}

$\mathbf{L \sim 1/\lambda_1}$ 

\vspace{2 mm}

Here the distribution of particle orientations varies in a nontrivial way with $x$: particles must travel a short distance from the boundary before their ``equilibrium" distribution $c + d (x - \cos \theta)$ is reached, which persists throughout the bulk of the channel. The statistics of this relaxation is embedded in the $L$ dependence of $\mathcal{A}(L)$ and $\mathcal{B}(L)$, and numerical calculation of the flux appears to be the best option. 

\vspace{2 mm}

$\mathbf{L \sim 1}$

\vspace{2 mm}

\begin{figure}
\centering
  \includegraphics[width=0.4347826087\linewidth,height=.3260869565\linewidth]{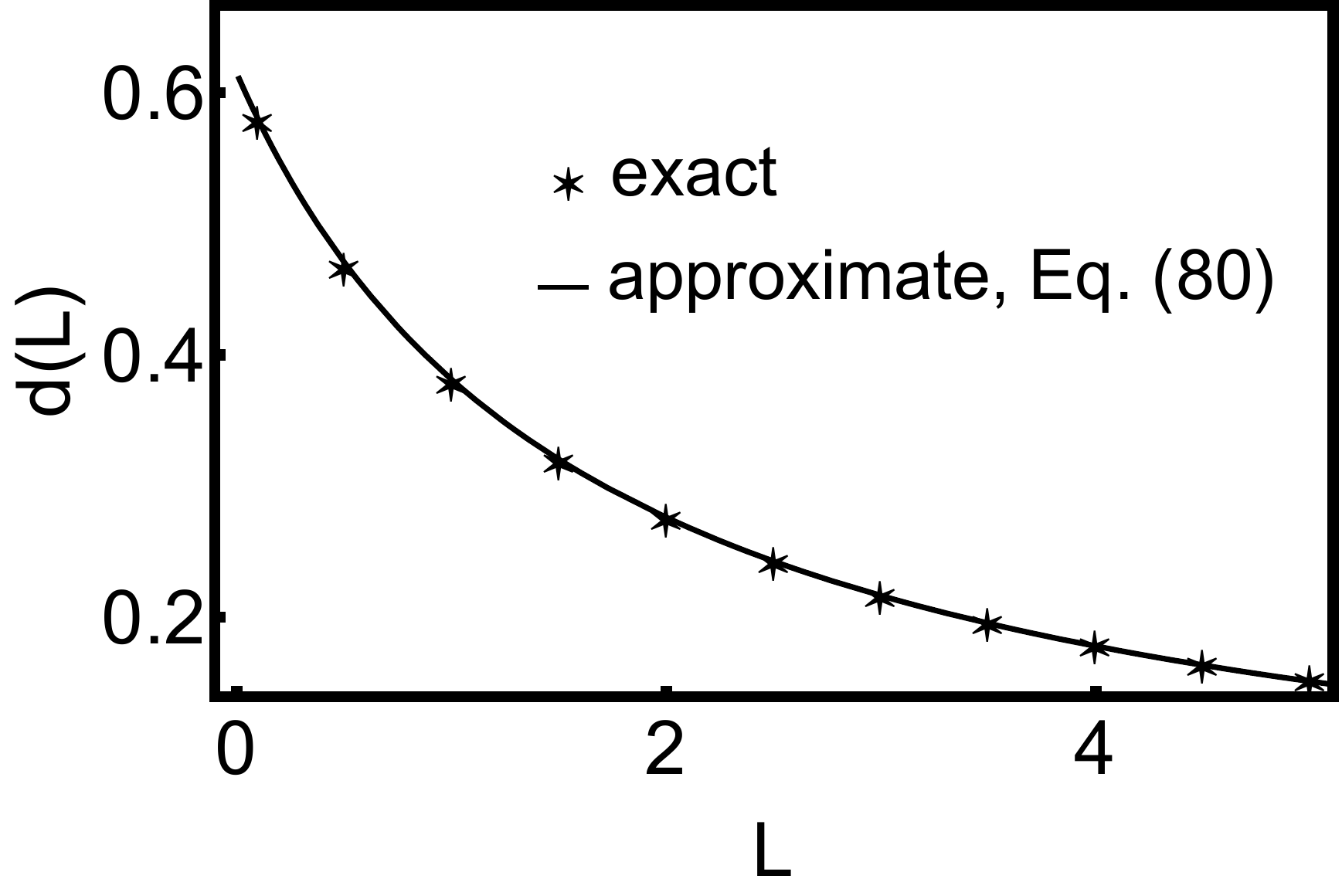}
\includegraphics[width=0.4347826087\linewidth,height=.3260869565\linewidth]{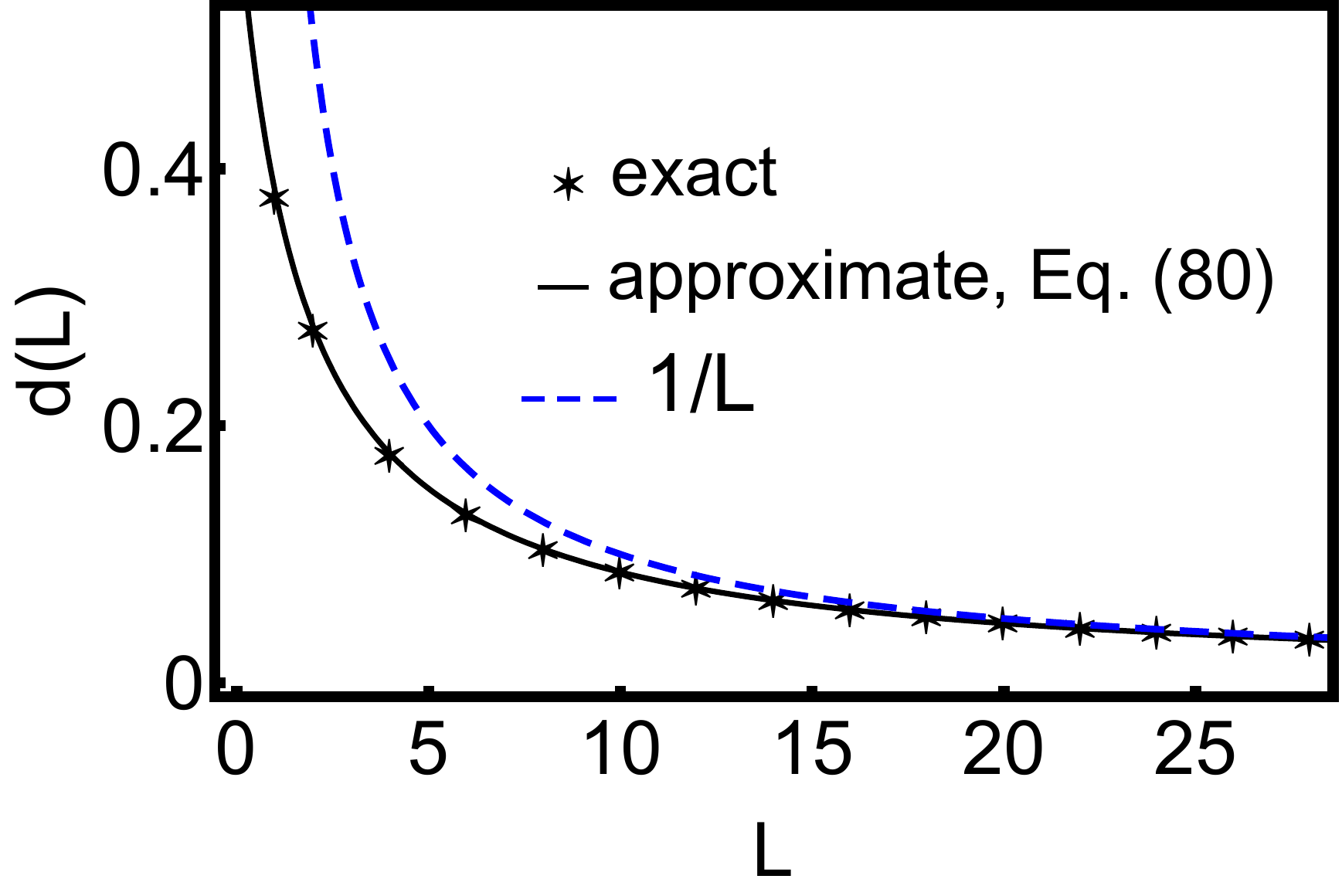}
  \caption{The expansion coefficient $d$ as a function of channel width $L$. Here $\rho_2 = 2$ and $\rho_1 = 1$. For reference, the black asterisks show the exact solution, obtained by evaluating the formal sum \eqref{sec3-formal} numerically to arbitrary precision. The solid curves show the approximations discussed in the main text.}
  \label{fig:beta-data}
\end{figure}

The intermediate range $L \sim 1$ can be interpreted in terms of the persistent motion of ABPs. Recall that in the non-dimensionalized model considered here, ABP trajectories are correlated over a distance equal to $1$. On the other hand, for distances much larger than $1$, trajectories decorrelate and the ABPs behave diffusively\cite{Marchetti2016}.  Thus, the region $L \sim 1$ interpolates between ballistic behavior at short distances and diffusive behavior at long distances. While the ballistic and diffusive limits are easy to understand, the intermediate length scale is mathematically nontrivial, and encodes much of the interesting behavior exhibited by ABPs (see e.g. Refs. \onlinecite{Wagner2017,Marchetti2016,Ni2015,Angelani2009}). The advantage of our technique is its ability to interpolate accurately in this regime, as illustrated in figure \ref{fig:beta-data}. 

\vspace{2 mm}

$\mathbf{L \gg 1}$

\vspace{2 mm}

In the large $L$ limit ABPs behave diffusively. Then we expect Fick's law of diffusion to hold, which in the present case says that the particle flux is $-D (\Delta \rho/L)$, where $D$ is the diffusivity. This prediction can be verified by expanding equation \eqref{d_large_L} for $L \gg 1$, which gives
\begin{equation}
\text{flux} = -\pi d \underset{L \gg 1}{\approx} -\frac{\pi \Delta \rho}{L},
\end{equation}
implying a diffusivity approximately equal to $\pi$. In figure \ref{fig:beta-data} the $1/L$ trend is compared with the exact solution, confirming its accuracy for $L \gg 1$.

\vspace{5mm}

\section{Summary} \label{sec-summary}

We have introduced a new method for solving two-way diffusion problems. The advantage of the technique is its ability to generate a sequence of approximate analytical solutions as terms in a systematic expansion. Convergence can be assessed using numerical estimates and holds for a variety of problems. We have illustrated these ideas in the context of a periodic problem and demonstrated the usefulness of the technique in obtaining physical insights.

\begin{acknowledgments}
C.G.W. acknowledges funding from the NSF (the Brandeis MRSEC DMR-1420382 and IGERT DGE-1068620). 
\end{acknowledgments}

\section*{Appendix A: Bounds on $||W_L||$} 

In this section we apply a WKB analysis to the case $A = -\partial^2 / \partial \theta^2$ and relate bounds on $||W_L|| = ||W(I-M_L)||$ to the eigenvalue spectrum. Our starting point is the easily checked identity
\begin{equation}
2||Wu||^2\=||u||^2-||u||_1^2-2\la P_+u,P_-u\ra.\label{starting}
\end{equation}
One can use this to obtain various formal bounds on $||W||$, e.g.
\begin{equation}
||W||^2 \leq \frac{1}{2} + \sup_{||u|| = 1} |\la P_+u,P_-u\ra|.
\end{equation}
Roughly speaking, the quantity  $|\la P_+u,P_-u\ra|$ expresses the ``overlap'' between the two half-range bases and thus estimates the relative size of the error in truncating the formal solution \eqref{sec3-formal}. As we now show, $|\la P_+u,P_-u\ra|$ also has a well-defined connection to the eigenvalue spectrum, which can be used to prove rigorous bounds on $||W||$. 


Since the analysis differs qualitatively depending on the algebraic multiplicity of $h(\theta)$ at the turning point, we consider first the general problem
\begin{equation}  
v''(\theta) = \lambda \theta^m v(\theta), \hspace{8mm} -T < \theta < T,
\label{eq:appendixA-general-turning-point}
\end{equation}
where $m$ is a positive, odd integer. Solutions of this are given by
\begin{equation}
v(\theta) = \left[\lambda^{1/(m+2)} \theta \right]^{1/2} J_{\pm \omega} \left( \frac{2i}{m+2} \lambda^{1/2} \theta^{(m+2)/2} \right),
\end{equation}
where $J_{\pm \omega}$ is a solution of Bessel's equation
\begin{equation}
x^2 J''_{\pm \omega} + x J'_{\pm \omega} + (x^2 - \omega^2) J_{\pm \omega} = 0, \hspace{5mm} \omega = 1 / (m+2).
\end{equation}
For convenience, define
\begin{align}
v_{-}(\theta) &=  \left[\lambda^{1/(m+2)} \theta \right]^{1/2} J_{-\omega} \left( \frac{2i}{m+2} \lambda^{1/2} \theta^{(m+2)/2} \right), \\
v_{+}(\theta) &=  \left[\lambda^{1/(m+2)} \theta \right]^{1/2} J_{\omega} \left( \frac{2i}{m+2} \lambda^{1/2} \theta^{(m+2)/2} \right).
\end{align}
These two functions form a basis of solutions to \eqref{eq:appendixA-general-turning-point}. Now, up to some constant factors which we take to be 1,
\begin{equation}
x^{\omega} J_{-\omega}(x) = 1 - x^2 + ... \ , \qquad x^{\omega} J_{\omega}(x) = x^{2 \omega} - x^{2 \omega + 2} + ... \ , 
\end{equation}
which leads to
\begin{equation}
v_{-}(\theta) = 1 + \lambda \theta^{m+2} + ... \ , \qquad v_{+}(\theta) = \lambda^{1/(m+2)} \theta + \left(\lambda^{1/(m+2)} \theta \right)^{1+(m+2)} + ... \ . \label{eq:vpm-small-theta}
\end{equation}
Assuming $m$ is odd, we expect oscillatory solutions where $\lambda \theta < 0$. In such a case the asymptotics of the $J_{\pm \omega}$ tell us that for large $|\lambda|$,
\begin{equation}
v_{\pm}(\theta) = |\lambda^{1/(m+2)} \theta|^{1/2} \cdot \frac{1}{|\lambda|^{1/4}|\theta|^{(m+2)/2}} \cdot \cos \left(\frac{2|\lambda \theta^{m+2}|^{1/2}}{m+2} - \frac{\pi}{4} \mp \frac{\pi}{m+2} \right). \label{eq:uplusminus}
\end{equation}

For any given boundary conditions on \eqref{eq:appendixA-general-turning-point}, each eigenfunction $v_{k}$ can be written as a linear combination of $v_{\pm}$: $v_{k} = a_k v_{+} +  b_k v_{-}$. We would like to scale these coefficients so that the normalization from the main text holds: $\int v_{k}^2 \theta^m d\theta = \sgn(k)$. Equation \eqref{eq:uplusminus} implies that to leading order in $|\lambda_k|$ the necessary scaling is $a_k \sim b_k \sim |\lambda_k|^{m/(4m + 8)}$. 

Now let us return to the quantity $\la P_+u,P_-u\ra$. Expanding $u = \sum a_k v_k$, this becomes
\begin{equation}
\la P_+u,P_-u\ra = \sum_{k>0} \sum_{j<0} a_j a_k \langle v_j, v_k \rangle. \label{eq:PplusPminus-expansion}
\end{equation}
Next, we relate the quantity $\langle v_j, v_k \rangle$ to the Wronskian of $v_j$ and $v_k$ evaluated at $\theta = 0$:
\beas
\l_j\la v_j, v_k\ra&=&\int_0^T\l_j h v_jv_k-\int^0_{-T}\l_jhv_jv_k
\=\int^T_0 v_j''v_k-\int^0_{-T}v_j''v_k\\
&=& 2(v_jv_k'-v_j'v_k)\big|_0+\l_k\la v_j,v_k\ra.
\eeas
In particular, if $\l_j$ and $\l_k$ have opposite signs we have  
\be\label{A1}
\big|\la v_j,v_k\ra\big|\=2\,\frac{|v_j(0)v'_k(0)-v_j'(0)v_k(0)|}
{|\l_j|+|\l_k|}.
\ee
Finally, in view of equation \eqref{eq:vpm-small-theta}, and using the appropriate normalization for $v_k$, we have that $v_j(0) \sim |\lambda_j|^{m/(4m + 8)}$ and $v'_j(0) \sim |\lambda_j|^{(4+m)/(4m + 8)}$. Thus,
\begin{align}
|\langle v_j, v_k \rangle| &\sim \frac{|\lambda_j|^{m/(4m + 8)}|\lambda_k|^{(4+m)/(4m + 8)} + |\lambda_j|^{(4+m)/(4m + 8)}|\lambda_k|^{m/(4m + 8)}}{|\l_j|+|\l_k|} \label{eq:innerproduct-scaling} \\
&\sim |\lambda_k|^{-(3m+4)/(4m+8)} \hspace{5mm} (j \text{ fixed}).
\end{align}
Using the same type of reasoning employed in a WKB approximation, this scaling in fact holds for any $h(\theta)$, with $m$ determined by the behavior of $h(\theta)$ near the turning point. With more careful attention to error terms and constant factors, one can then combine this result with equations \eqref{starting} and \eqref{eq:PplusPminus-expansion} to obtain various concrete bounds on $||W||$.

Unfortunately, we were not able to show in this way that $||P|| \cdot ||W|| < 1$ for the periodic problem \eqref{example1} -- \eqref{example3}, although this bound is consistent with numerics (see appendix B). Nevertheless, two conclusions of general relevance come out of this analysis. First, in view of \eqref{eq:innerproduct-scaling}, small but nonzero eigenvalues may be responsible for large bounds on $||W||$. This conclusion is confirmed by numerics, and motivated the generalized formulation of the problem in section \ref{sec-alternate-formulation}. Second, the asymptotics on $|\langle v_j, v_k \rangle|$ show faster decay  as the algebraic multiplicity $m$ of the turning point increases. Thus, problems with higher order turning points may have smaller $||W||$, consistent with the numerics in appendix B (the case $h(\theta) = \sgn(\theta)$ being construed as having an algebraic multiplicity of zero).


%
%
%
%
%
%
%
%
%
%
%
%
%

\section*{Appendix B: Calculation of $||W_{L, N}||$}

As in the main text, we consider the finite-dimensional space $\mathscr{H}_{1,\leq N}=\mathrm{span}\{v_{j}\}_{\left\vert
j\right\vert \leq N}$, within which a function $u\left( \theta \right) 
$ can be expanded:%
\begin{equation}
u\left( \theta \right) =\sum_{j=1}^{N}\left( a_{j}v_{j}+b_{j}v_{-j}\right) .
\end{equation}
Thus, $u\left( \theta \right) $ can also be written as a column vector $%
\overrightarrow{u}=(a_{1},a_{2},\ldots ,a_{N},b_{1},b_{2},\ldots ,b_{N})^{T}$%
. Now, $\left\Vert
W_{L, N}\right\Vert $ can be estimated as follows. First, for any $\overrightarrow{u} \in $ $\mathscr{H}_{1,\leq N}$, there exist $2N\times 2N$ matrices $A$ and $S
$ such that%
\begin{eqnarray}
\left\Vert \overrightarrow{u} \right\Vert^2  &=&\overrightarrow{u}^{T}A%
\overrightarrow{u}; \\
\left\Vert W_{L, N} \overrightarrow{u} \right\Vert^2  &=&\overrightarrow{u}%
^{T}S\overrightarrow{u}.
\end{eqnarray}

\bigskip These matrices can be constructed in terms of the various inner
products of the $v_{j}$. We want to estimate 
\begin{equation}
\left\Vert W_{L, N}\right\Vert^2 =\mathrm{\sup_{\left\Vert \overrightarrow{u} \right\Vert =1}}\left\Vert
W_{L, N}\overrightarrow{u} \right\Vert^2. 
\end{equation}

Such an extremum can be calculated from the Lagrangian%
\begin{eqnarray}
f(\overrightarrow{u},\lambda ) &=&\left\Vert W_{L, N}\overrightarrow{u}
\right\Vert^2 -\lambda \left( 1-\left\Vert\overrightarrow{u} \right\Vert^2
\right)  \\
&=&\overrightarrow{u}^{T}S\overrightarrow{u}+\lambda \left( 1-%
\overrightarrow{u}^{T}A\overrightarrow{u}\right) .
\end{eqnarray}

Differentiating with respect to $\overrightarrow{u}$, and using the fact
that $S$ and $A$ are symmetric matrices, we get%
\begin{equation}
\frac{\partial f(\overrightarrow{u},\lambda )}{\partial \overrightarrow{u}}%
=2S\overrightarrow{u}-2\lambda A\overrightarrow{u}.
\end{equation}

Thus, extrema satisfy the generalized eigenvalue problem%
\begin{equation}
S\overrightarrow{u}=\lambda A\overrightarrow{u}\label{1}
\end{equation}

In particular, multipliying from the left by $\overrightarrow{u}^{T}$ gives%
\begin{equation}
\left\Vert W_{L, N}\overrightarrow{u} \right\Vert^2 =\lambda .
\end{equation}

Since $A$ and $S$ are positive definite, the absolute maximum of $\left\Vert
W_{L, N}\overrightarrow{u} \right\Vert^2$ subject to $\left\Vert \overrightarrow{u} \right\Vert =1$ is just the largest eigenvalue of
the problem (\ref{1}). 

Values of $\left\Vert W_{L, N}\right\Vert^2$ are calculated in this way in Maple as a function of $N$ and fit to the power law Ansatz $A_0 - B_0 N^{-\nu}$ using the function `NonlinearFit' with initial value $\nu = 1$. Ranges of $N$ used in each fit are (1000, 6000) for the step problem, (1000, 2800) for the linear and periodic problems, and (2000, 8000) for the cubic problem. Values of the fit parameters are shown in table \ref{table:power-law-fits}, while the fits themselves are plotted in figure \ref{fig:w_estimates} in the main text. 

Note that our analysis assumes that the $L$ dependence of $W_{L, N}$ is negligible, i.e. that terms of order $e^{-|\lambda^*| L}$ can be ignored, where $\lambda^*$ is the smallest eigenvalue in magnitude. Including this $L$ dependence only modifies the estimates downwards.

\begin{table}[H]
\caption{Power law fits $A_0 - B_0 N^{-\nu}$ to $||W_{L, N}||^2$.} 
\centering
\begin{tabular}{p{2cm}      p{3cm}	p{2cm}	p{2cm}	p{2cm}} 
\hline\hline 
$h(\theta)$ & $A_0$ & $B_0$ & $\nu$ \\ [1ex]
\hline 
$\sgn(\theta)$ & 5.33 & 6.33 & $0.149$  \\ 
$\theta$ & 0.916 & 1.02 & 0.194  \\
$\cos(\theta)$ & 0.884  & 1.20 & 0.223 \\
$\theta^3$ & 0.187 & 0.182 & 0.209 \\ [1ex] 
\hline 
\end{tabular}
\label{table:power-law-fits}
\end{table} 

\section*{Appendix C: Periodic problem} 

Here we present the details of the calculations for the periodic problem in section \ref{sec-periodic-problem}.

\vspace{5 mm}

\textbf{Eigenvalues and eigenfunctions}

\vspace{2 mm}

Numerical calculations of the spectrum are carried out as in Ref. \onlinecite{Wagner2017}. There the eigenfunctions are expressed as even/odd Fourier series of period $2 \pi$. This Ansatz leads to a second-order recurrence for the Fourier coefficients, which can be written as the eigenvalue equation of a tridiagonal matrix. This matrix equation can be solved numerically in a computer algebra system (here we use Maple), allowing accurate computation of the first several hundred eigenvalues and eigenfunctions. 

\vspace{5 mm}

\textbf{Estimate on $||P||$} 

\vspace{2 mm}

Let $\mathscr{H}_1^\perp$ be the orthogonal complement of $\mathscr{H}_1$ with respect to $\la\,,\,\ra$. Recall also that $\mathscr{H}_0 = \text{span}\{1,g_L\}$. Then $P \mathscr{H}_1$ = $\mathscr{H}_1$ and $P \mathscr{H}_0 = 0$. Writing an element $v\in \mathscr{H}$ as $v_1+v^\perp$, $v_1\in \mathscr{H}_1$ and $v^\perp\in \mathscr{H}_1^\perp$,
we have
$$
Pv\= v_1+Pv^\perp.
$$
Suppose $||v||=1$.
For a given $w=v^\perp\ne0$, the maximum of $||Pv||$ occurs when $v_1$ is taken
to be a positive multiple of $Pw$, $v_1=\sqrt{1-||w||^2}/||Pw||\cdot Pw$, 
giving
$$
||Pv||\=\sqrt{1-||w||^2}+||Pw||.
$$
Thus we want to find 
\be\label{rho}
\rho\ =\sup_{w\perp \mathscr{H}_1,||w||=1}||Pw||
\ee
and then maximize
$$
\sqrt{1-s^2}+\rho s,\qquad 0\le s\le 1.
$$
The maximum occurs where  $s=\rho/\sqrt{1+\rho^2}$ and is 
\be\label{P-norm}
||P||\=\sqrt{1-\frac{\rho^2}{1+\rho^2}}+\rho\cdot\frac{\rho}{\sqrt{1+\rho^2}}\=\sqrt{1+\rho^2}.
\ee
It remains to determine $\rho$.

Here $g(\th)=-\cos\th$.
It is convenient to replace $g_L$ with $\wt g_L=g_L-\frac12L$, so that
\begin{equation}
\wt g_L(\th)\=\left\{\begin{matrix} -\cos(\th)-\tfrac12L &\ \ {\rm if}\ \ \cos\th>0;\\
-\cos(\th)+\tfrac12L&\ \ {\rm if}\ \ \cos\th < 0.\end{matrix}\right.
\end{equation}
Thus $\wt g_L$ is odd with respect to $\sgn\cos\th$.
Using this and the identity $4\cos^3\th=\cos 3\th+3\cos\th$, one calculates
\be
\la 1,1\ra\=4,\qquad\la 1,\wt g_L\ra\=0,\qquad\la \wt g_L,\wt g_L\ra\=
\frac83+\pi L+L^2.
\ee
Thus we can normalize and take $e_1,e_2$ as orthonormal basis
for $\mathscr{H}_0$, where
\be
e_1(\th)\=\frac12,\qquad e_2(\th)\=\frac1\s_1 \wt g_L,\quad\s_1=\sqrt{\frac83+\pi L+L^2}.
\ee
It follows from \eqref{1g} that the functions $\sgn\cos\th$ and $|\cos\th|$ are 
a basis for $\mathscr{H}_1^\perp$.  We obtain an orthonormal basis $f_1,f_2$ by setting
\begin{equation}  
f_1(\th)\=\frac12\sgn \cos\th, \qquad f_2\=\frac1\s_2|\cos\th|,\ \ \s_2=\sqrt{\frac83}.
\end{equation}
Then 
\beas
&&\la e_1,f_1\ra\=0\=\la e_2,f_2\ra;\\
&&\la e_1,f_2\ra\= \frac\pi{2\s_2}\ \equiv\ \frac1{r_1},\quad\la e_2,f_1\ra\=
-\frac{2L+\pi}{2 \sigma_1}\ \equiv\ \frac1{r_2}.
\eeas
It follows that
\beas
\mathscr{H}_1\ \ni\ f_1-\frac1{\la e_2,f_1\ra}\cdot e_2\=
f_1-r_2 e_2;\\
\mathscr{H}_1\ \ni\ f_2-\frac1{\la e_1,f_2\ra}\cdot e_1\=
f_2-r_1 e_1.
\eeas
Applying $P$, we obtain
$$
Pf_1\=f_1-r_2 e_2;\qquad Pf_2\=f_2-r_1 e_1.
$$
Thus for $w=\a f_1+\beta f_2\in \mathscr{H}_1^\perp$,
\beas
||Pw||^2&=&||w-\a r_2 e_2-2\b r_1e_1)||^2\\
&=&\a^2(r_2^2-1)+\b^2(r_1^2-1).
\eeas
It can be checked that  $r_1^2> r_2^2$, so the maximum for $\a^2+\b^2=1$ is 
$\r=\sqrt{r_1^2-1}$
and \eqref{P-norm} gives
$$
||P||\=\sqrt{\r^2+1}\=r_1\=\frac2\pi\,\sqrt{\frac83}
\=\frac{4\sqrt6}{3\pi}\=1.0395...\ .
$$

\vspace{5mm}

\textbf{Expansion coefficients}

\vspace{3mm}

At zeroth order, the coefficients $c_0$ and $d_0$ are determined from equation \eqref{sec4-coeff1}:
\begin{equation}
\int \left(w - c_0 - d_0 g_L\right) h d\theta = 0 = \int \left(w - c_0 - d_0 g_L\right)ghd\theta,
\label{appendix-coeff1}
\end{equation}
which gives
\begin{align}
c_0 &= \frac{\rho_1+\rho_2}{2} - \frac{L}{2L + \pi} \Delta \rho; \\
d_0 &= \frac{2}{2L+\pi} \Delta \rho.
\end{align}
We get $a_j^0$ from equation \eqref{sec4-coeff2}:
\begin{align}
a_j^0 &= (-\Delta \rho + d_0 L) X_j; \\
X_j &=\text{sgn}(\lambda_j) \int_{\cos \theta > 0} v_j \cos \theta d\theta,
\end{align}
where we have used the fact that $\int_{\cos \theta > 0} v_j \cos \theta d\theta = -\int_{\cos \theta < 0} v_j \cos \theta d\theta$.

At first order, we find
\begin{align}
c_1 &= \frac{L}{2L+\pi} \mathcal{A}(L) \left[\Delta \rho - d_0 L \right]; \\
d_1 &= -\frac{2}{2L+\pi} \mathcal{A}(L) \left[\Delta \rho - d_0 L \right]; \\
\mathcal{A}(L) &\equiv  \sum_{k<0} X_k^2 \left(1-e^{\lambda_k L} \right).
\end{align}
In simplifying these expressions, we have used the fact that in our normalization, $v_j(\theta + \pi) = v_{-j}(\theta)$. 

The result for $a_j^1$ is more complicated, and involves the introduction of an additional quantity $\mathcal{C}_j(L)$:
\begin{align}
a_j^1 &= (-\Delta \rho + d_0 L) \mathcal{C}_j(L) + d_1 L X_j; \\
\mathcal{C}_j(L) &\equiv \text{sgn}(\lambda_j) \sum_{k<0} X_k \left(1-e^{\lambda_k L} \right) \int (Q_+ v_k - Q_- v_{-k}) v_j \cos \theta d\theta.
\end{align}
Finally, $d_2$ takes the form
\begin{align}
d_2 &= \left(\frac{2}{2L+\pi}\right) \mathcal{B}(L) (\Delta \rho - d_0 L) + L \left(\frac{2}{2L+\pi}\right) d_1 \mathcal{A}(L); \\
\mathcal{B}(L) &= -\sum_{k<0} \mathcal{C}_k X_k \left(1-e^{\lambda_k L} \right).
\end{align}

\begin{figure}
\centering
  \includegraphics[width=0.47\linewidth,height=.3260869565\linewidth]{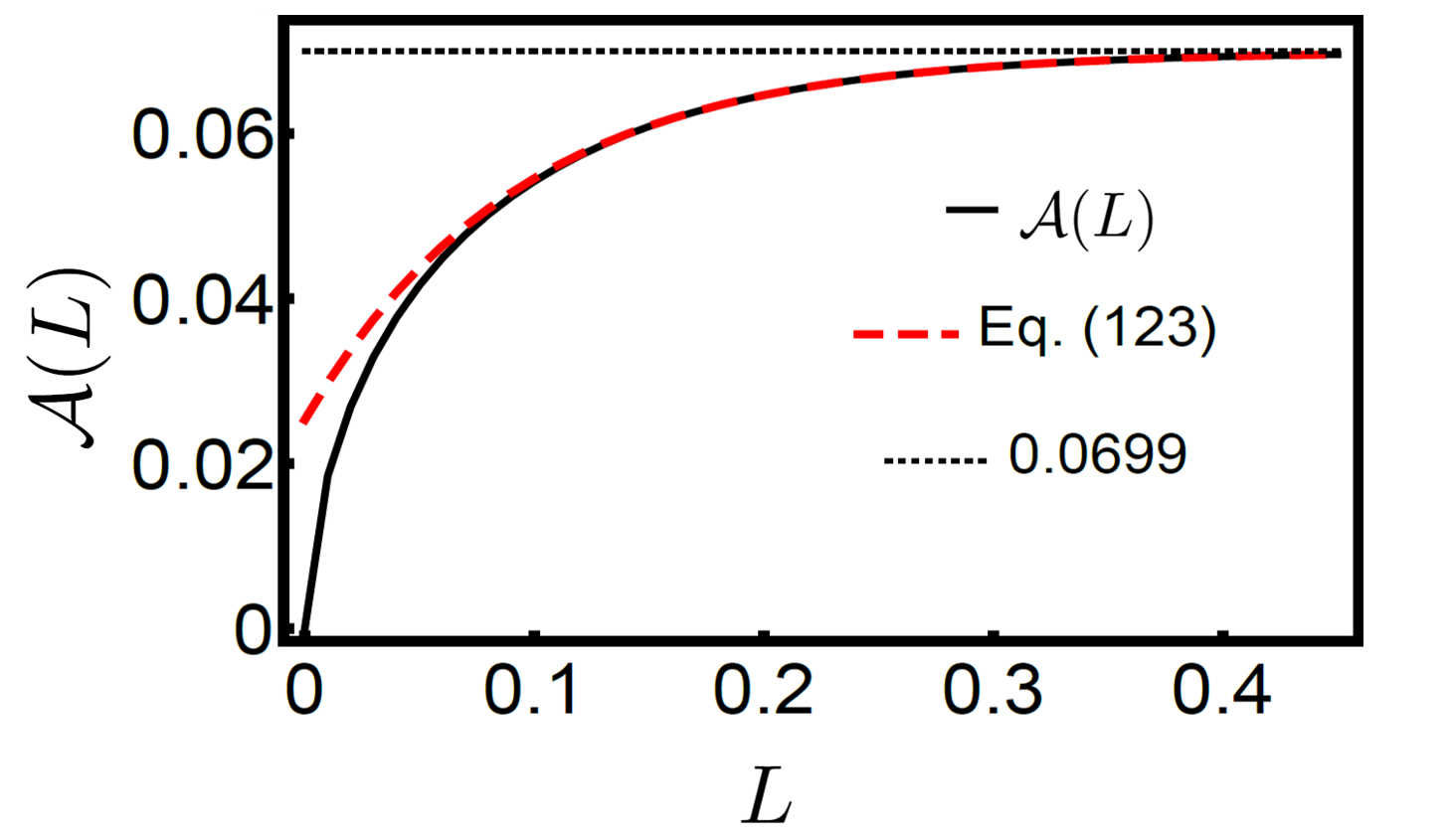}
\includegraphics[width=0.477\linewidth,height=.3260869565\linewidth]{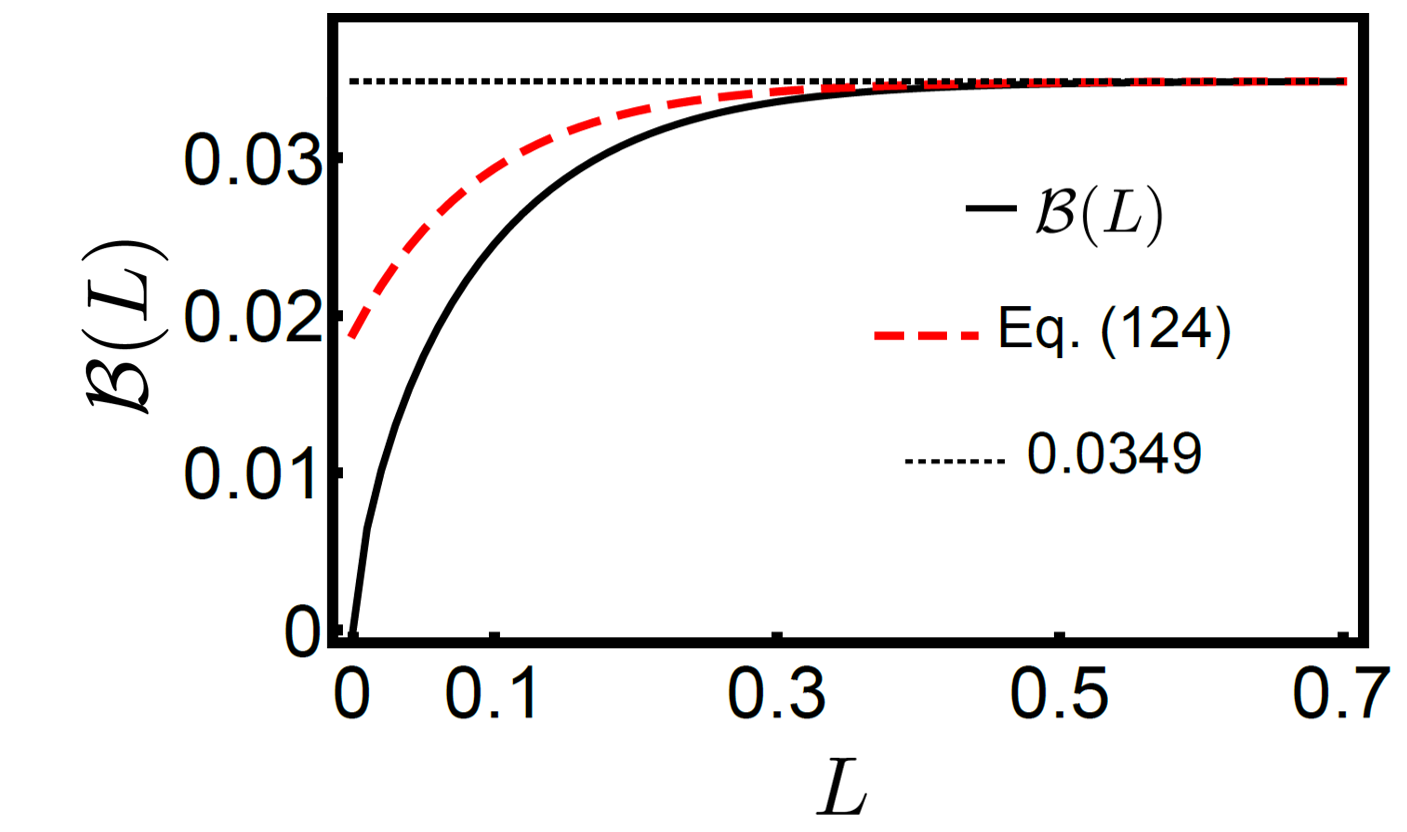}
  \caption{The quantities $\mathcal{A}(L)$ and $\mathcal{B}(L)$ occurring in the expressions for the expansion coefficients. Their behavior for arbitrary $L$, shown in black, is nontrivial. However, as $L$ grows larger than $1/\lambda_1 \simeq 0.1$ they rapidly approach constant values (dashed black lines). Shown in red are the slightly improved expressions, equations \eqref{appendixA-approx1} and \eqref{appendixA-approx2}, which retain terms of order $e^{-\lambda_1 L}$.}
  \label{fig:AandBcoefficients}
\end{figure}

The complexity of the solution is embedded in the nontrivial dependence of $\mathcal{A}(L)$, $\mathcal{B}(L)$, and $\mathcal{C}_j(L)$ on $L$. However, as discussed in the main text, substantial simplifications can be made for $L \gg 1/\lambda_1$, in which case factors such as $e^{-|\lambda_k| L}$ can be dropped. For instance, retaining only $e^{-\lambda_{1} L}$ and dropping all other exponentials leads to
\begin{align}
\mathcal{A}(L) &\approx 0.0699 - 0.0446 e^{-\lambda_{1} L} \label{appendixA-approx1} \\
\mathcal{B}(L) &\approx 0.0349 - 0.016 e^{-\lambda_{1} L} \label{appendixA-approx2}
\end{align}
As shown in figure \ref{fig:AandBcoefficients}, these are good approximations for sufficiently large $L$.

\bibliographystyle{apsrev4-1}
\bibliography{bib}

\end{document}